\documentclass[%
 reprint,
nofootinbib,
 amsmath,amssymb,
 aps,
]{revtex4-2}

\usepackage{graphicx}
\usepackage{dcolumn}
\usepackage{bm}
\usepackage[hidelinks,colorlinks=true,allcolors=blue]{hyperref}

\def\vev#1{{\langle{#1}\rangle}}

\usepackage{color}
\usepackage{slashed}
\usepackage[utf8]{inputenc}
\usepackage[compat=1.1.0]{tikz-feynman}
\usepackage{amsmath,amssymb,multirow,graphicx, subfigure}
\usepackage[margin=0.5in]{geometry}

\def\tr{\text{tr}}

\def\vev#1{{\langle{#1}\rangle}}

\def\hat{\widehat}
\def\bar{\overline}

\def\g{{\gamma}}

\def\e{{\epsilon}}

\def\l{{\lambda}}
\def\L{{\Lambda}}
\def\m{{\mu}}

\begin{document}

\preprint{APS/123-QED}

\title{Beta Functions of 2d Adjoint QCD}

\author{Aleksey Cherman}
\email{acherman@umn.edu}
\author{Maria Neuzil}%
\email{neuzi008@umn.edu}
\affiliation{%
 School of Physics and Astronomy, University of Minnesota, Minneapolis MN 55455, USA
}%


\date{\today}

\begin{abstract}
    We discuss the long-distance physics of 2d adjoint QCD when it is viewed as an
     effective field theory, and determine the $\beta$ functions for its
     two classically-marginal four-fermi operators.  These four-fermion
     terms preserve the invertible symmetries of the kinetic terms, and
     they have important implications at long distances if they are
     generated at short distances.  Our results are likely to be
     important for future numerical lattice Monte Carlo studies of 2d
     adjoint QCD.            
\end{abstract}
            
            \maketitle

\section{Introduction}

One-flavor massless $SU(N)$ adjoint QCD is a popular setting for
exploring the physics of confinement; see e.g. Refs.~
\cite{Dalley:1992yy,Kutasov:1993gq,Bhanot:1993xp,Lenz:1994du,Gross:1997mx,
Davies:1999uw,Armoni:2003gp,Armoni:2003fb,Armoni:2003yv,Armoni:2004uu,
Unsal:2007jx,Kovtun:2007py,Poppitz:2012sw,Poppitz:2012nz,Cherman:2016jtu,
Komargodski:2017keh,Gomis:2017ixy, Bergner:2018unx,Poppitz:2021cxe,Anber:2022qsz,Cherman:2019hbq,Komargodski:2020mxz}.
This theory --- which we will call QCD[Adj] --- has just two fields, an $SU(N)$
gauge field $a_{\mu}$ and a massless adjoint-representation Majorana
fermion $\psi$, but its physics is very rich.   First, the fact that $\psi$ is in the adjoint representation
means that there is a $\mathbb{Z}_N$ $1$-form symmetry that acts on
Wilson loops, which ensures that confinement is a well-defined notion; see e.g. Refs.~\cite{Gross:1980br,Gaiotto:2014kfa}.  Second,
massless QCD[Adj] enjoys $\mathcal{N}=1$ supersymmetry in
$d=4$ spacetime dimensions, but it is not supersymmetric when $d<4$.  Third, QCD[Adj] has a rich matrix-like large $N$ limit in $d \ge 2$, in
contrast to $SU(N)$ QCD with fundamental fermions, where the standard large $N$
limit becomes vector-like in $d=2$.    Finally, the large $N$ limit of QCD[Adj] is believed to be equivalent to the large $N$ limit of QCD
with a Dirac fermion in the two-index anti-symmetric `AS'
representation\cite{Armoni:2004uu}.  At $N = 3$ the AS representation is
the same as the (anti-)fundamental representation, so studying QCD[Adj] yields lessons for an unusual but phenomenologically viable\cite{Armoni:2004uu,Cherman:2009fh,Buchoff:2009za,Cohen:2009wm,Lebed:2010um,Cherman:2012eg} large $N$ limit of $SU(3)$ QCD in
any $d\ge2$.  All of these features make QCD[Adj]
a very interesting setting for exploring the physics of confinement.

The Euclidean action of QCD[Adj] is
\begin{align}
    S = \int d^{d}x \left[\frac{1}{2g^2} \tr f_{\mu \nu}^2 + \tr \bar{\psi} \slashed{D} \psi + \ldots \right]\,.
    \label{eq:action_ldots}
\end{align}
The ellipsis represents other local terms built out of $a_{\mu}$ and
$\psi$ that are consistent with the symmetries of the kinetic terms.   In addition to the $\mathbb{Z}_N$ $1$-form symmetry, the invertible
symmetries include fermion parity $(-1)^F$, charge conjugation $C$, time
reversal $T$, a coordinate reflection symmetry $R$, and a discrete
chiral symmetry in even spacetime dimensions: $\mathbb{Z}_2$ in $d=2$
and $\mathbb{Z}_N$ in $d=4$.

In this paper, we will follow a common Wilsonian perspective and interpret Eq.~\eqref{eq:action_ldots} as an effective field theory, so that Eq.~\eqref{eq:action_ldots} is viewed as the longer-distance effective action corresponding to some short-distance description involving additional
heavy degrees of freedom.  For instance, we could add an extra scalar field $\phi$  of mass
$M \gg g$ with a Yukawa interaction of the form $ \int d^{d} x \,
y \,\phi\, \tr\, \bar{\psi}\psi $ to Eq.~\eqref{eq:action_ldots}, and then consider the physics for energies $\ll M$. As another example, Eq.~\eqref{eq:action_ldots} could be the coarse-grained effective description of a short-distance spacetime lattice model with lattice spacing $a$. 
We will try to understand the
`universal' aspects of the long-distance physics which are independent of the details
of such short-distance modifications, provided that the invertible symmetries of Eq.~\eqref{eq:action_ldots} mentioned above
are not explicitly broken at short distances.  The dependence of the
physics on various possible symmetry-preserving short-distance modifications is parameterized by
symmetry-preserving $\ldots$ terms in Eq.~\eqref{eq:action_ldots}.

In $d \ge 3$, any symmetry-preserving terms other than the kinetic terms in Eq.~\eqref{eq:action_ldots}
are technically irrelevant.   The only apparent exception is the $d=4$ term
$\frac{i\theta}{32\pi^2} \int d^4x\, \epsilon^{\mu\nu \alpha \beta } \tr
f_{\mu \nu} f_{\alpha \beta}$ with $\theta = \pi$, but this term has no
physical effects since it can be absorbed in the phase of the massless
fermion field. Therefore, if one's goal is to understand the universal
aspects of the long-distance physics of QCD[Adj], then in $d \ge 3$ it is safe to delete the ellipsis
in Eq.~\eqref{eq:action_ldots}. 

In this paper we focus on the physics in $d=2$, where the situation is
different thanks to the fact that there are two classically-marginal
four-fermion terms consistent with all the invertible symmetries
mentioned above.  As a result the EFT action of 2d QCD[Adj] can be
written as\cite{Cherman:2019hbq}
\begin{align}
    S = \int &d^{2}x \bigg[\frac{1}{2g^2} \tr f_{\mu \nu}^2 + \tr \bar{\psi} \slashed{D} \psi  \nonumber\\
    + &\frac{1}{2N}\l_j \tr [\bar{\psi}  \g^{\mu} \psi \bar{\psi} \g_{\mu} \psi]  - \frac{1}{N^2}\l_m \tr [\bar{\psi} \psi] \tr[\bar{\psi} \psi] \bigg]\,.
    \label{eq:action_with_four_fermi}
\end{align}
The $\lambda_j$ term is like the Thirring interaction for a Dirac
fermion\cite{Thirring:1958in}, while the $\lambda_m$ term is a version of the
Gross-Neveu interaction\cite{Gross:1974jv}. As we review below, these two terms are identical at $N=2$.

A very interesting feature of 2d QCD[Adj] is that when $\lambda_m$ is turned off, the model has an exotic `non-invertible' symmetry\cite{Komargodski:2020mxz}; see also \cite{Delmastro:2021otj,Delmastro:2022prj}.   This exotic symmetry would be explicitly broken by the adjoint quark mass term $m \tr \bar{\psi} \psi$, which would also break the $\mathbb{Z}_2$ chiral symmetry.   The $\lambda_m$ term is the square of the mass term and does not break chiral symmetry, but it turns out that it \emph{does} break the non-invertible $0$-form symmetry\cite{Komargodski:2020mxz}.  The other four-fermion term, $\l_j \tr [\bar{\psi}  \g^{\mu} \psi \bar{\psi} \g_{\mu} \psi]$, does not break any of the symmetries.  Due to the infamous subtleties involved in defining chiral lattice fermions (reviewed in e.g. Ref.~\cite{Kaplan:2009yg}), the non-invertible symmetry discovered in Ref.~\cite{Komargodski:2020mxz} seems unlikely to be preserved in spacetime lattice formulations of 2d QCD[Adj]. One might therefore worry that without fine-tuning, numerical lattice simulations of 2d QCD[Adj] might end up reaching continuum limits described by Eq.~\eqref{eq:action_with_four_fermi} that  have non-zero values of $\lambda_m$ and $\lambda_j$.  This is one of several motivations for understanding the impact of the four-fermion terms on the long-distance physics.

We calculate the $\beta$ functions of $\lambda_j$ and $\lambda_m$ in Section~\ref{sec:beta_functions} after setting out our conventions in Section~\ref{sec:conventions}.  We find that
with a natural choice of signs for $\lambda_j$ and $\lambda_m$, these couplings generically become large in the long distance limit.  We then discuss the implications of the RG flow for confinement in
Section~\ref{sec:IR_results}.  Finally, readers may have noticed that with $\lambda_m = \lambda_j = 0$, 2d QCD[Adj] becomes super-renormalizable, and it may seem impossible for a four-fermion term to be generated through radiative corrections in such a QFT.  While this is true in the naive continuum theory, in Section~\ref{sec:lattice} we will show how it can be false on the lattice. We then conclude in Section~\ref{sec:conclusions}.

\section{Conventions}
\label{sec:conventions}

For concreteness, we use the following representation of 2d Euclidean gamma matrices,
\begin{align}\label{eq:gamma_matrices}
    \gamma^1=
    \begin{pmatrix}
        0 & 1 \\
        1 & 0
    \end{pmatrix} \quad
    \gamma^2=
    \begin{pmatrix}
        1 & 0 \\
        0 & -1
    \end{pmatrix} \,,
\end{align}
and define $\g\equiv i\gamma^1 \gamma^2$.
The charge conjugation matrix $C=-\gamma$ satisfies
 $C\g^{\mu} C^{-1}=-(\g^{\mu})^T$ 
and  we define the charge conjugate of $\psi$ to be
$\psi^c=C^{-1}\bar{\psi}^T$; see e.g. Ref.~\cite{Stone:2020vva} for an extensive discussion of possible conventions. The Majorana condition $\psi^c=\psi$ then implies $\bar{\psi}=\psi^T C^T=\psi^T \g $.
The Euclidean action takes the form
\begin{align}\label{eq:full_action}
    S=\int d^2 x \,\, \bigg[& \frac{N}{2\lambda} \tr[f_{\mu\nu} f^{\mu\nu}] +\tr [\bar{\psi}\slashed{D}\psi]+m\,\tr [\bar{\psi} \psi] \\
    + \frac{1}{2N}\l_j &\tr [\bar{\psi}  \g^{\mu} \psi \bar{\psi} \g_{\mu} \psi] - \frac{1}{N^2}\l_m \tr [\bar{\psi} \psi] \tr[\bar{\psi} \psi]  \nonumber \bigg]
\end{align}
where $\psi = \psi_A t^A$ is a Majorana fermion in the adjoint representation of $SU(N)$, $t^A$ are the generators of $SU(N)$ in the fundamental representation with $A = 1, \ldots, N^2-1$, the trace is taken over color indices with $\tr \, t_A t_B = \frac{1}{2} \delta_{AB}$, the transpose in $\bar{\psi}$ acts on spinor (but not color) indices, $\lambda = g^2 N >0$ is the 't Hooft coupling, and $N \ge 2$. We
are ultimately interested in the $m=0$ theory, but we will use the mass term to regulate low-momentum divergences in the Feynman diagram calculations. 

The $\tr [\bar{\psi} \psi] \tr[\bar{\psi} \psi]$ term is an adjoint Majorana version of the four-fermi interaction in the Gross-Neveu model\cite{Gross:1974jv}.  Our sign of the $\lambda_m$ term in Eq.~\eqref{eq:full_action} matches that of Gross and Neveu, who fixed the sign of the four-fermion coupling in their model by requiring it to match the effective coupling that would be induced by integrating out a heavy scalar $\phi$ with a Yukawa coupling to the mass operator, which takes the form $ \int d^{2} x \,
y \,\phi\, \tr\, \bar{\psi}\psi $ in QCD[Adj].  We also chose the sign of the $\tr [\bar{\psi}  \g^{\mu} \psi \bar{\psi} \g_{\mu} \psi]$ term to match the standard sign of the four-fermion coupling in the Thirring model\cite{Thirring:1958in}.\footnote{Our sign choices are also compatible with the sign choices of Ref.~\cite{Bedaque:1991xr} which studied the $\beta$ functions of classically marginal couplings, including some four-fermion couplings, in generalized Schwinger models with $N$ fermions.  We also note that Ref.~\cite{Onder:2023tei} recently calculated the $\beta$ functions of four-fermion deformations of some 2d chiral gauge theories, while Ref.~\cite{Cherman:2022ecu} studied four-fermion couplings in the one-flavor charge $N$ Schwinger model.}   Happily, these two choices are nicely correlated with each other in QCD[Adj] thanks to the fact that for $N=2$\cite{Cherman:2019hbq}
\begin{align}\label{eq:N=2_tr_reln}
    \tr [\bar{\psi}  \g^{\mu} \psi \bar{\psi} \g_{\mu} \psi] = - \tr[\bar{\psi}\psi]\tr[\bar{\psi} \psi] \,.
\end{align}
This also means that for $N=2$ there is only one independent symmetry-preserving four-fermion operator.  

Before moving on we should note that it is sometimes possible for four-fermion terms to have the `wrong' sign without physical problems.  This is famously the case for the Thirring coupling $g_t$ for a single Dirac fermion, which can be negative as long as $g_t > - \pi/2$.  While we will focus on $\lambda_j \ge 0$ and $ \lambda_m \ge 0$, it would be interesting to understand whether there are well-motivated UV completions of 2d adjoint QCD that yield negative values for these couplings.

Finally, we note that the fermion $\psi$ can be decomposed into right- and left-moving components, $R$ and $L$, which in this basis takes the form 
\begin{align}
    \psi=\frac{1}{\sqrt{2}}\left(R\begin{pmatrix}
        1 \\ i
    \end{pmatrix}+L\begin{pmatrix}
        i \\ 1
    \end{pmatrix}\right) \,.
\end{align}
This leads to a useful alternate form of the four-fermi interactions,
\begin{align}
     \tr[\bar{\psi} \g^{\mu}\psi \bar{\psi} \g^{\mu} \psi]&= 4 \,\tr[LLRR]\nonumber \\
     \tr[\bar{\psi} \psi]\tr[\bar{\psi} \psi]&=-4\,\tr[LR]\tr[LR]\,.
\end{align}

\section{Calculating beta functions}
\label{sec:beta_functions}
Our goal is to calculate the $\beta$ functions of $\lambda_j$ and $\lambda_m$ to one loop accuracy.  The gauge coupling is dimensionful while $\lambda_j,\lambda_m$ are dimensionless, so gluons cannot contribute to the the $\beta$ functions of $\lambda_j$ and $\lambda_m$ at any finite loop order.  Therefore we do not consider gluon loops in this section.

We calculate the $\beta$ function in two different ways:  from the operator product expansion (OPE) (see e.g. Ref.~\cite{cardy_scaling_book}), and from standard considerations of one-loop Feynman diagrams.  The agreement of the results provides a check on our calculations.

\subsection{Beta functions from OPEs}
Let us define the operators
\begin{align}
    \mathcal{O}_j&= \frac{2}{N} :\tr[LLRR]: \nonumber \\
    \mathcal{O}_m &= \frac{4}{N^2} :\tr[LR]\tr[LR]:
\end{align}
where $:\,:$ denotes normal ordering. The $\beta$ functions for $\l_j$ and $\l_m$ are then given by\cite{cardy_scaling_book}\footnote{We thank Diego Delmastro for introducing us to this method.}
\begin{align}\label{eq:generic_beta_OPE}
    \beta_j&=\pi (c^j_{jj}\l_j^2+2c^j_{mj}\l_m \l_j+c^j_{mm}\l_m^2) \nonumber
    \\
    \beta_m&=\pi (c^m_{jj}\l_j^2+2c^m_{mj}\l_m \l_j+c^m_{mm}\l_m^2)
\end{align}
where $c_{IJ}^K$ are the OPE coefficients
\begin{align}
    \mathcal{O}_I(z)\mathcal{O}_J(0)=\sum_K\frac{c^K_{IJ}\mathcal{O}_K(0)}{z^{\Delta_I+\Delta_J-\Delta_K}}+\dots
\end{align}
where $I, J, K = j, m$ and here $\Delta_{I} = 2$ for all $I$. 

When $m=0$, the propagators are
\begin{align}
    \vev{{L^a}_{b}(z){L^{c}}_{d}(0)}&=\frac{1}{4\pi i z}\left(\delta^a_{d}\delta^c_{b}-\frac{1}{N}\delta^a_{b}\delta^c_{d}\right)\nonumber \\
    &=-\vev{{R^a}_{b}(z){R^c}_{d}(0)}\,,
\end{align}
where $a,b,c,d=1,\dots,N$ are fundamental color indices.  To determine the OPE coefficients we use Wick's theorem to evaluate contractions in products of $\mathcal{O}_j$ and $\mathcal{O}_m$ and find 
\begin{align}
    \mathcal{O}_j(z)\mathcal{O}_j(0)&=\frac{1}{z^2}\left(-\frac{1}{4\pi^2}\mathcal{O}_j(0)\right) +\dots\nonumber \\
    \mathcal{O}_m(z)\mathcal{O}_j(0)&=\frac{1}{z^2}\left(\frac{1}{N^2 \pi^2}\mathcal{O}_j(0)-\frac{1}{2 \pi^2}\mathcal{O}_m(0)\right) +\dots\nonumber \\ \mathcal{O}_m(z)\mathcal{O}_m(0)&=\frac{1}{z^2}\left( -\frac{1-3/N^2}{\pi^2}\mathcal{O}_m(0)\right) +\dots \,.
\end{align}
Note that $\mathcal{O}_{m}$ does not appear in the $1/z^2$ part of the contraction of $\mathcal{O}_j(z)\mathcal{O}_j(0)$, and vice versa.  Applying Eq.~\eqref{eq:generic_beta_OPE} we find
\begin{align} \label{eq:b_fns}
    \beta_j&=-\frac{\l_j}{\pi}\left(\frac{1}{4}\l_j-\frac{2}{N^2}\l_m\right) \nonumber \\
    \beta_m&=-\frac{\l_m}{\pi}\left( \left( 1-\frac{3}{N^2}\right)\l_m +\l_j\right)\,,
\end{align}
and observe that there is neither a $\lambda_m^2$ term in $\beta_j$ nor a $\lambda_j^2$ term in $\beta_m$.

\subsection{Beta functions from Feynman diagrams}

\begin{figure}[h!]
\includegraphics[width=0.77\columnwidth,angle=-90]{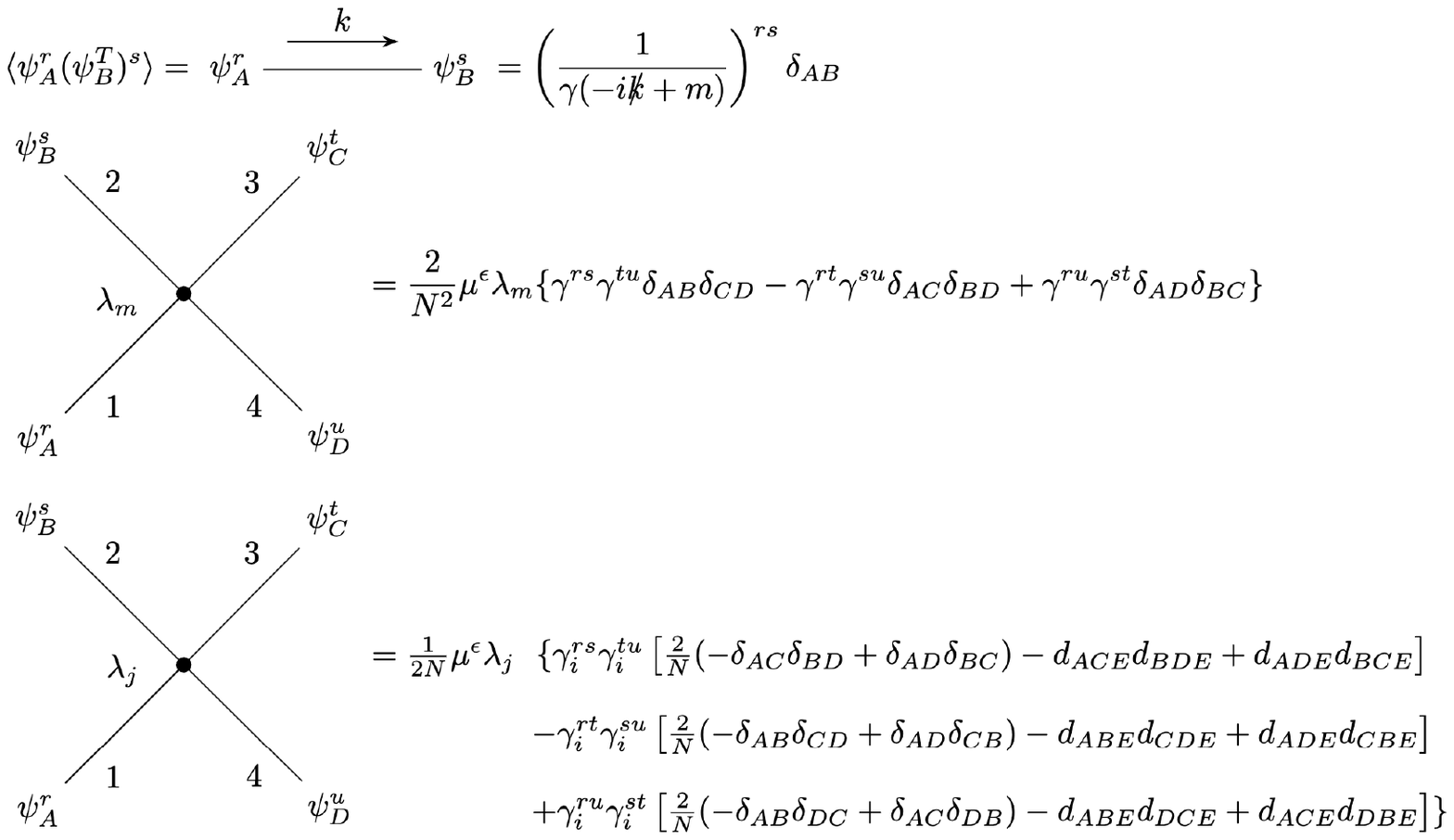}
\caption{Feynman rules for the action in Eq.~\eqref{eq:S_counterterms}. Here $A,B,C,D,E=1,\dots,N^2-1$ are adjoint color indices, $r,s,t,u$ are spinor indices, $i$ is a Lorentz index, $d_{ABC}$ are the totally-symmetric structure constants of $SU(N)$, and summation over repeated indices is implied.}
\label{fig:feynman_rules}
\end{figure}

We check Eq.~\eqref{eq:b_fns} by calculating $\beta_j$ and $\beta_m$ using standard Feynman diagram methods with the Feynman rules in Fig.~\ref{fig:feynman_rules}.  Dropping the gauge field (since it does not contribute to our calculation), we express the action with counterterms in dimension $d=2-\e$ as
\begin{align}\label{eq:S_counterterms}
    S&=\int d^d x \,\, \frac{1}{2}\psi_{A}^T \g( \slashed{\partial}+m)\psi_{A} +\frac{1}{2}\psi_{A}^T \g( \delta_{\psi}\slashed{\partial}+\delta_m) \psi_{A}  \nonumber \\&- \frac{1}{2N}\m^\e \l_{j} \tr [\psi^T  \g^{i} \psi \psi^T  \g_{i} \psi] - \frac{1}{N^2}\m^\e \l_{m} \tr [\psi^T \g \psi] \tr[\psi^T \g \psi] \nonumber \\
    &- \frac{1}{2N}\m^\e\delta_{\l_j} \tr [\psi^T  \g^{i} \psi \psi^T  \g_{i} \psi] - \frac{1}{N^2}\m^\e\delta_{\l_m}\tr [\psi^T \g \psi] \tr[\psi^T \g \psi]\,,
\end{align}
where $A=1,\dots,N^2-1$ is an adjoint color index and the mass scale $\mu$ has been introduced so that $\l_j$ and $\l_m$ are dimensionless. We include the fermion mass to regulate IR divergences and take $m\to0$ and the end. 
The $\beta$ functions can be built from
\begin{align}
    \L_j(\l_j,\l_m)\equiv \l_j\frac{Z_{\l_j}}{Z^2_{\psi}}\qquad \L_m(\l_j,\l_m)\equiv \l_m\frac{Z_{\l_m}}{Z^2_{\psi}}
\end{align}
where the $Z$'s are related to the $\delta$'s by
\begin{align}
    \delta_{\psi}=Z_{\psi}-1 \,,\,\, \delta_{\l_{j,m}}=\l_{j,m}(Z_{\l_{j,m}}-1) \,.
\end{align}
The $\beta$ functions can then be obtained from the system of equations (see e.g. \cite{Zinn-Justin:2002ecy}) 
\begin{align}
    0&=\epsilon \L_j+\beta_j\frac{\partial \L_j}{\partial \l_j}+\beta_m\frac{\partial \L_j}{\partial \l_m} \nonumber \\
    0&=\epsilon \L_m+\beta_j\frac{\partial \L_m}{\partial \l_j}+\beta_m\frac{\partial \L_m}{\partial \l_m}\,.
\end{align}

\begin{figure}[h]
\centering
\includegraphics[width=0.4\columnwidth,angle=-90]{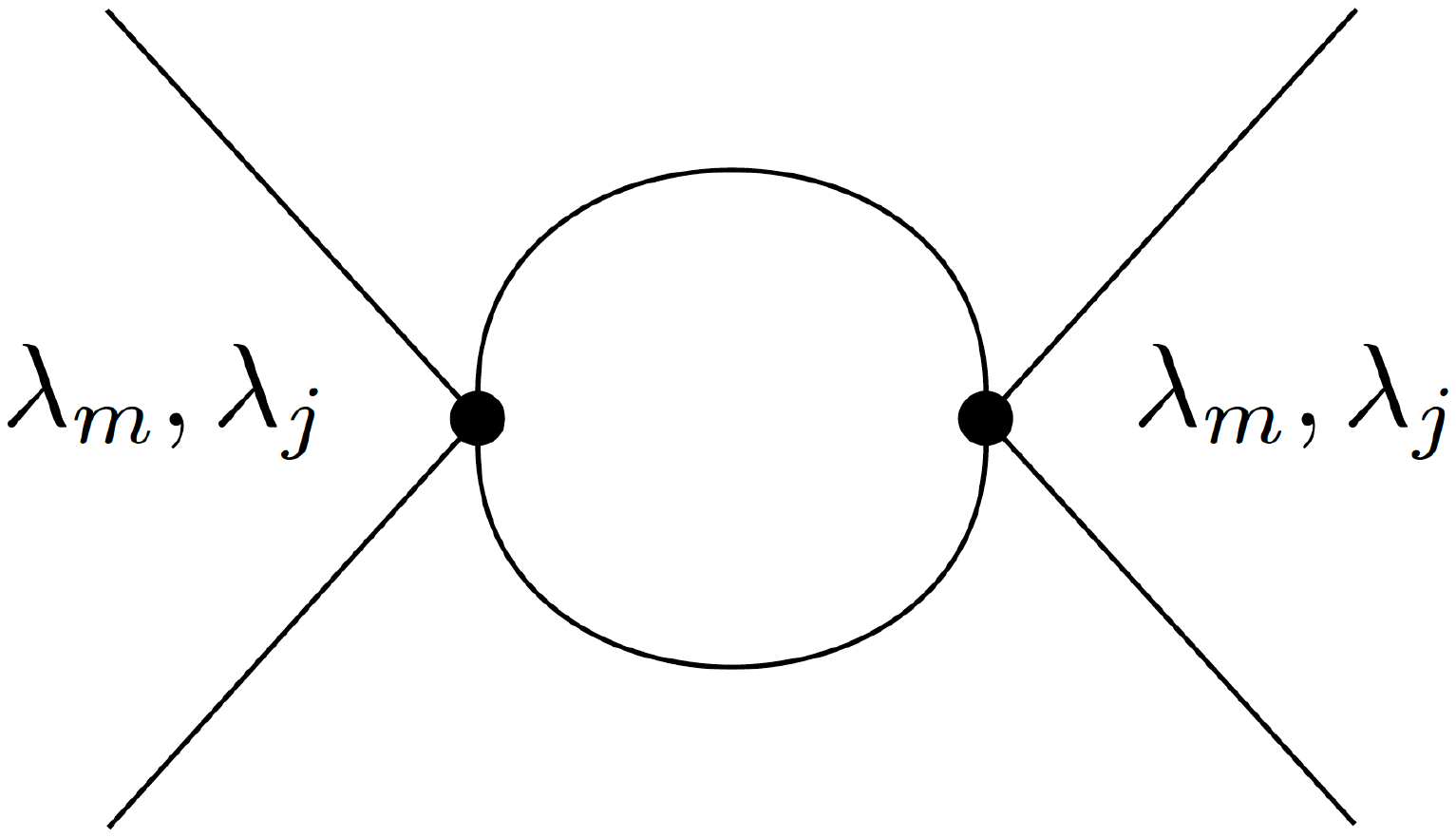}
\caption{The one-loop Feynman diagrams needed for our calculation of the $\beta$ functions all take the basic form sketched in this diagram. 
}
\label{fig:loop_diagram}
\end{figure}

We evaluate bubble diagrams involving $\lambda_m$ and $\lambda_j$ of the form shown in Fig.~\ref{fig:loop_diagram} using the Feynman rules in Fig.~\ref{fig:feynman_rules}, and simplify the results using standard $SU(N)$ identities (see e.g. \cite{Haber:2019sgz}) to obtain 
\begin{align}\label{eq:counter_terms}
    \delta_{\psi}&=0 \nonumber \\
    \delta_{\l_j}&=-\frac{1}{4\pi \e}\l_j^2  +\frac{2}{N^2\pi \e}\l_j\l_m \nonumber \\
    \delta_{\l_m}&=-\frac{1-3/N^2}{\pi \e}\l_m^2 -\frac{1}{\pi \e}\l_j\l_m\,.
\end{align}

The resulting $\beta$ functions exactly match Eq.~\eqref{eq:b_fns}.  As another consistency check, we note that when $N =2$, $\beta_j$ and $\beta_m$ coincide, in the sense that if we turn off e.g. $\lambda_m$ when calculating $\beta_j$ and vice versa, we get the same result.  This is consistent with the fact that when $N=2$ the $\mathcal{O}_j$ and $\mathcal{O}_m$ operators are identical.

\section{The long-distance limit}
\label{sec:IR_results}
We now discuss the implications of our results for the long-distance physics of 2d QCD[Adj].  We first discuss the behavior of the couplings in the long-distance limit in Section~\ref{sec:RG_flow}, and then summarize the consequences for confinement in Section~\ref{sec:confinement}.

\subsection{Renormalization group flow}
\label{sec:RG_flow}

We first observe that  the $\beta$ functions in Eq.~\eqref{eq:b_fns} say that if at the short distance cutoff we have $\lambda_j>0$ but $\lambda_m = 0$, then $\lambda_m$ remains zero in the IR as long as the one-loop approximation can be trusted, and vice versa.  The fact that turning on $\lambda_j$ does not generate $\lambda_m$ is expected from symmetry considerations, because the theory with $\lambda_m = 0$ has an enhanced (non-invertible) symmetry\cite{Komargodski:2020mxz}.  But the fact that turning on $\lambda_m$ does not induce the $\lambda_j$ coupling cannot be explained by any known symmetry, so this seems likely to be a one-loop artifact. 

For $N=2$ there is only one independent four-fermion coupling consistent with the invertible symmetries of 2d QCD[Adj], and it is asymptotically free.  It flows to strong coupling at long distances, so if it is generated with a small positive coefficient at short distances, it will have important quantitative effects in the IR.   Whether it also has important qualitative effects is tied up with the anomaly structure of the theory, as we will discuss below.  

\begin{figure}[h]
    \centering
    \includegraphics[width=0.9\columnwidth]{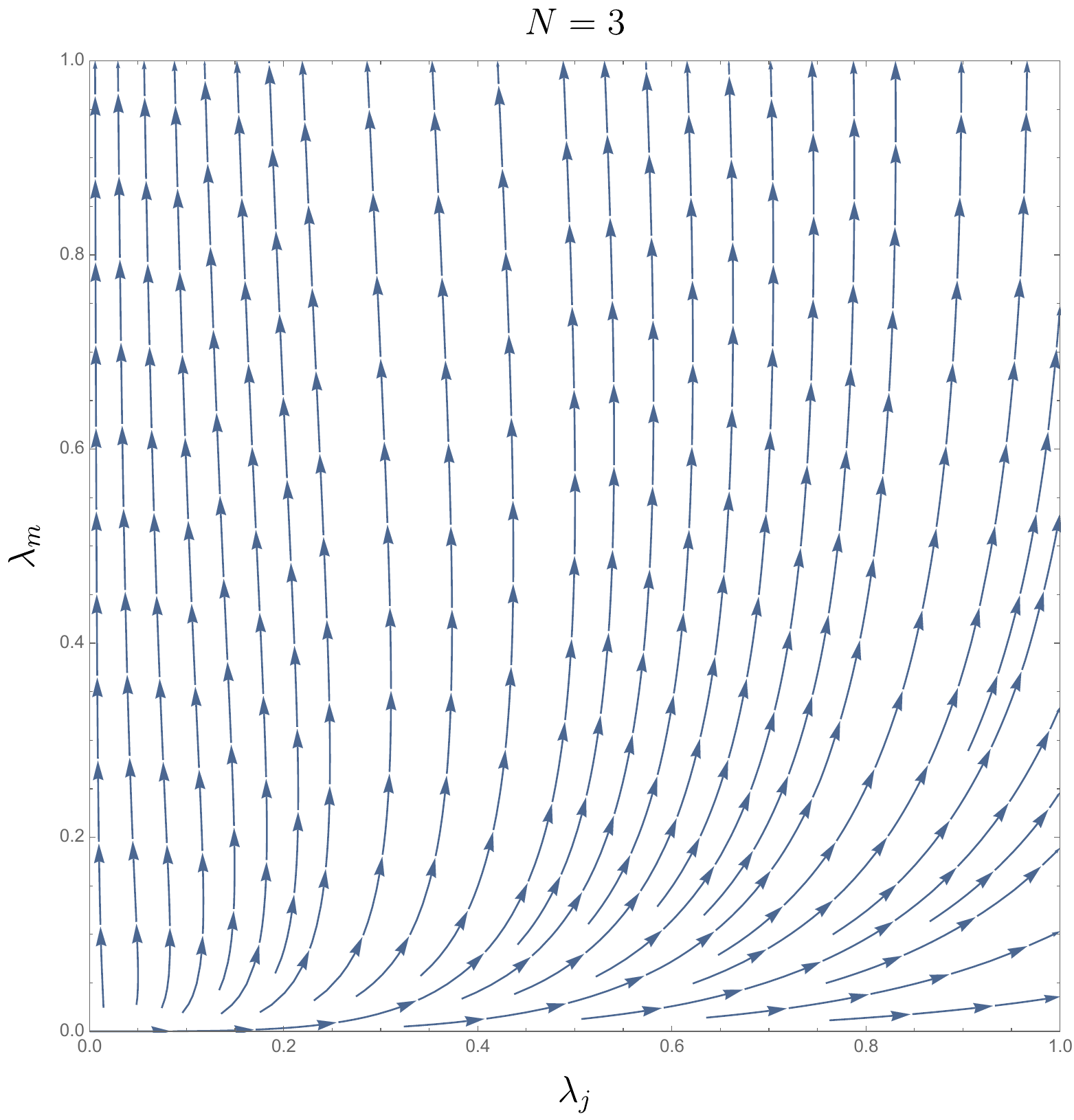}
    \caption{One-loop renormalization group flow of the four-fermion couplings $\lambda_m, \lambda_j$ of 2d $N=3$ QCD[Adj], with the arrows pointing toward the long-distance limit.   The $\lambda_m$ coupling always grows as we flow to the long-distance limit, while the behavior of $\lambda_j$ depends on the starting point in parameter space. While the plot suggests that  $\l_j$ flows to zero if $\l_j \ll \l_m$, this conclusion is not reliable because $\lambda_m$ becomes large in the same limit, leading to a breakdown of the one-loop approximation. 
    }
    \label{fig:RG_flow}
\end{figure}
    
The RG flow of the $\l_j,\l_m$ couplings for $N=3$ is illustrated in Fig.~\ref{fig:RG_flow}.   As one would expect from a continuum analysis of a super-renormalizable QFT, if $\lambda_m = \lambda_j = 0$ at the short-distance cutoff, then these couplings remain zero as one goes to long distances.    However, if $\lambda_m > 0$ at short distances, then it increases in the long-distance limit regardless of the value of $\lambda_j$.  The behavior of $\l_j$ depends on the relative sizes of $\l_j$ and $\l_m$ at the short-distance cutoff.  If $\lambda_j$ is sufficiently large compared to $\lambda_m$ at the cutoff, then both $\lambda_j$ and $\lambda_m$ increase as we flow to long distances.  If $\lambda_j$ is much smaller than $\lambda_m$ at the cutoff, then $\lambda_j$ decreases as we flow to long distances,   while $\lambda_m$ increases.  However,  the one-loop approximation breaks down once either coupling becomes large, and so at finite $N$ a perturbative calculation is not sufficient to determine the long-distance fate of $\lambda_j$ when $0<\lambda_j \ll \lambda_m$ at the short-distance cutoff.  

In the large $N$ limit the situation simplifies: the double-trace coupling $\lambda_m$ does not affect the RG flow of the single-trace coupling $\lambda_j$. Therefore in the large $N$ limit our one-loop analysis allows us to conclude that \emph{both} $\lambda_j$ and $\lambda_m$ flow to strong coupling in the long distance limit, and it is natural to expect $1/N$ corrections to be quantitavely small already at $N=3$ in this model with adjoint matter, where the large $N$ expansion goes in powers of $1/N^2$. 

We can summarize our discussion of the RG flow by saying that if some small positive $\lambda_j, \lambda_m$ four-fermion couplings are induced from short-distance effects, then they become large in the long-distance limit, and must be taken into account to understand the long distance physics.  

\subsection{Consequences for confinement}
\label{sec:confinement}

To understand the qualitative consequences of the discussion above we have to review the 't Hooft anomalies of 2d QCD[Adj].  We first recall that the lowest critical dimension for a discrete $1$-form symmetry is $d=3$\cite{Gaiotto:2014kfa}, so it is generally natural to expect 2d gauge theories with $\mathbb{Z}_N$ symmetries to confine test quarks since their $1$-form symmetry cannot be spontaneously broken; see Ref.~\cite{Nguyen:2024ikq} for a beautiful effective field theory explanation, as well as Ref.~\cite{Cherman:2021nox} for an earlier closely-related discussion.  However, there is an important subtlety: in $d=2$, if a $\mathbb{Z}_N$ $1$-form symmetry has a mixed 't Hooft anomaly with a $0$-form symmetry, then both symmetries will be at least partially spontaneously broken on $\mathbb{R}^2$, with some subtle features like the presence of `universes'\cite{Hellerman:2006zs,Komargodski:2020mxz}.  This is nicely illustrated by the behavior of the charge $N$ Schwinger model\cite{Anber:2018jdf,Komargodski:2020mxz,Cherman:2022ecu}, which is in a fully deconfined phase on $\mathbb{R}^2$ because the rank of its $\mathbb{Z}_N$ $0$-form chiral symmetry precisely matches the rank of its $\mathbb{Z}_N$ $1$-form symmetry, and these two symmetries have a mixed 't Hooft anomaly.   

Since 2d QCD[Adj] only has a $\mathbb{Z}_2$ chiral symmetry, the invertible symmetries leave no room for this sort of anomaly-driven complete deconfinement for $N>2$. (If $N$ is even there is a mixed chiral/$1$-form anomaly that leads to Wilson loops in representations with $N$-ality $N/2$ to be deconfined\cite{Cherman:2019hbq}.)  However, as we already mentioned, 2d QCD[Adj] has an exotic non-invertible symmetry so long as $\lambda_m = 0$ and $m=0$\cite{Komargodski:2020mxz}. The topological line operators generating this non-invertible symmetry carry all possible charges under the $\mathbb{Z}_N$ $1$-form symmetry, and this leads to the same type of anomaly-driven complete deconfinement seen in the charge $N$ Schwinger model.

Komargodski et. al. observed that turning on a small $\lambda_m$ coupling produces a non-vanishing string tension\cite{Komargodski:2020mxz} proportional to $\lambda_m$.  However, if $\lambda_m$ were marginally irrelevant, this $\lambda_m$-induced string tension would vanish in the long-distance limit.  But here we have shown that $\lambda_m$ is marginally \emph{relevant}.  This  implies that if a tiny $\l_m >0 $ is induced from some short-distance regularization, then it will grow at long distances, badly breaking the non-invertible symmetry and inducing confinement with a string tension set by some combination of the 't Hooft coupling $\lambda$ and the strong scale $\Lambda_m$ of $\lambda_m$.  Similarly, if a small $\lambda_j>0$ is induced in the action of 2d QCD[Adj] from some short-distance physics, we expect that e.g. the particle masses will be sensitive to the strong scale $\Lambda_j$ associated to $\lambda_j$, in addition to their sensitivity to the dimensionful 't Hooft coupling $\lambda$.

In summary, the four-fermion interactions of 2d QCD[Adj] can have important quantitative and qualitative effects on the long-distance physics, for example driving massless 2d QCD[Adj] to confine fundamental test quarks for any $N>2$.  If $\lambda_j$ and $\lambda_m$ are induced in the effective action of 2d QCD[Adj] from some short-distance physics, they must be carefully taken into account when studying the long-distance correlation functions of the theory.

\section{Four-fermion operators and the lattice}
\label{sec:lattice}

There are several ways to quantitatively explore the long-distance physics of 2d QCD[Adj].  One approach is to work on the light-cone, using either discretized light-cone quantization (DLCQ)\cite{Dalley:1992yy,Kutasov:1993gq,Bhanot:1993xp,Gross:1997mx,Dempsey:2022uie,Demeterfi:1993rs,Dempsey:2021xpf,Dempsey:2022uie,Trittmann:2023dar} or  light-cone conformal truncation\cite{Katz:2013qua,Anand:2020gnn}.   These methods have the advantage that they appear to preserve the non-invertible $0$-form symmetry of 2d QCD[Adj]; see the discussion in e.g. Ref.~\cite{Kutasov:1994xq} regarding the decoupling of left- and right-handed excitations in light of the results of Ref.~\cite{Komargodski:2020mxz}.  The light-cone approaches also have the nice feature of being relatively inexpensive numerically, at least at large $N$. They also have some disadvantages, ranging from challenges with directly studying spontaneous symmetry breaking using light-cone methods --- see e.g.~Refs.~\cite{Robertson:1992nj,Bender:1992yd,Pinsky:1993yi,Pinsky:1994si} --- as well as difficulties with calculating correlation functions of large Wilson loops, which are the most natural observables for studying the physics of confinement and the realization of the $\mathbb{Z}_N$ $1$-form symmetry.

Another approach, which is especially widely used to study gauge theories in $d>2$, is to perform numerical Monte Carlo calculations of Euclidean correlation functions using lattice gauge theory.  So far the only lattice gauge theory calculations of 2d adjoint one-flavor QCD in the massless limit is Ref.~\cite{Dempsey:2023fvm}, which constructed a remarkable Hamiltonian lattice discretization that correctly captures all of the invertible symmetries of the model, including chiral symmetry.  However, the numerical analysis of Ref.~\cite{Dempsey:2023fvm} focused on $N=2$, where (a) $\lambda_m$ is equivalent to $\lambda_j$, and (b) the model is deconfined simply due to the anomalies of the invertible symmetries.  The continuum-limit behavior of $N>2$ QCD[Adj] defined on spacetime (or for that matter spatial) lattices is not yet fully clear.  

As discussed above, if we set $\lambda_m = \lambda_j = 0$ in the classical Lagrangian of the continuum theory, then these couplings  stay zero in the quantum theory by dimensional analysis.  But it is less clear what would happen to $\lambda_j$ and $\lambda_m$ if we start with a lattice discretization. The lattice brings in another dimensionful parameter, the lattice spacing $a$, and necessarily breaks some spacetime symmetries, such as translation symmetry. It also often breaks or modifies some internal symmetries, such as chiral symmetry\cite{Kaplan:2009yg}.  Can four-fermion interaction terms that cannot be generated in a continuum field theory be radiatively generated when such a theory is discretized? 

A naive massless lattice fermion action leads to $2^d$ massless `doubler' fermions in the continuum limit; see e.g. \cite{Wilson:1974sk,Karsten:1980wd,Kaplan:2009yg}.  There are several known ways around this, but all of them do something subtle to chiral symmetry.  For example, let us consider Wilson lattice fermions\cite{Wilson:1974sk}.  The idea of Wilson fermions is to add a (dangereously) irrelevant term to the fermion action which explicitly breaks chiral symmetry and also breaks the degeneracy between the fermion doubler modes, so that the continuum action resulting from coarse-graining the lattice action becomes (in a continuum notation)
\begin{align}\label{eq:wilson_action}
S_W = \int d^{d}x &\bigg[\frac{1}{2g^2} \tr f_{\mu \nu}^2 + \tr \bar{\psi} \slashed{D} \psi   \nonumber \\
&+ m \tr \bar{\psi}\psi + r a \tr \bar{\psi} \slashed{D}^2 \psi + \ldots \bigg] \,,
\end{align}
where the $r$ term is the Wilson term, $a$ is the lattice spacing, and $\ldots$ stands for other terms induced by coarse-graining the lattice action.  For generic values of $m$, one gets $2^d$ heavy fermions with mass $m \sim 1/a$ in the continuum limit.  However, it is known that one can tune the bare quark mass $m$ to get a single light or massless physical quark in the continuum limit, while the extra $2^{d}-1$ doubler quarks remain heavy, with masses $\sim 1/a$.

\begin{figure}[h!]
\centering
\includegraphics[width=0.4\columnwidth,angle=-90]{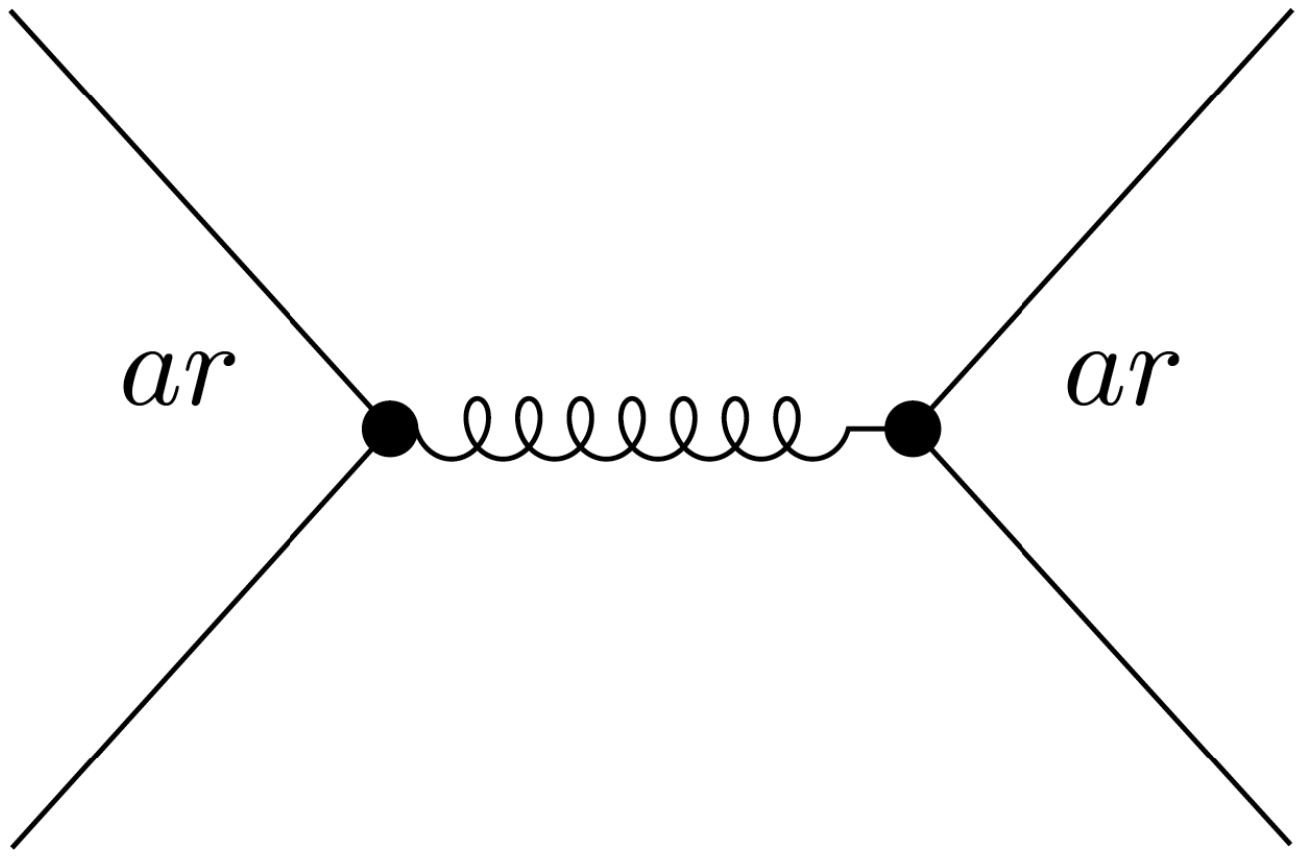}
\caption{A tree-level gluon-exchange diagram from the Wilson term in Eq.~\eqref{eq:wilson_action}, which leads to an effective $\l_j\sim r^2g^2a^2$.
}
\label{fig:gluon_exchange}
\end{figure}

It is easy to verify that the extra terms hiding in $\ldots$ include four-fermion interactions. In particular it is easy to check that the \emph{tree-level} gluon exchange diagram in e.g. Fig.~\ref{fig:gluon_exchange} produces an effective $\lambda_j \sim r^2 g^2 a^2$ interaction.  This should not be surprising because no symmetry forbids such an interaction, and the extra dimensionful scale $a$ invalidates the naive continuum argument for the impossibility of gluon loops producing classically-marginal local four-fermion interactions.  It is natural to expect that higher-order diagrams induced by the Wilson term also produce the $\lambda_m$ coupling, since $\lambda_m$ is also not forbidden by either symmetries or dimensional analysis arguments.  While these effective four-fermion couplings $\lambda_j,\lambda_m$ will necessarily appear suppressed by positive powers of $g a$, and hence appear to be small near the continuum limit where $g a \ll 1$, these couplings run and can become large at long distances.  Tuning $m$ to reach the chiral limit clearly does not generically tune four-fermion couplings to zero.  Therefore if one were to formulate 2d QCD[Adj] on the lattice using Wilson fermions, one would study the physics of Eq.~\eqref{eq:action_with_four_fermi} with non-zero four-fermion couplings $\lambda_j, \lambda_m$.  To avoid this, one would have to turn on $\lambda_j, \lambda_m$ in the bare lattice action and fine-tune them.

To show that Wilson fermions are not special in inducing four-fermion interactions in lattice field theories --- even though it appears to be impossible to induce such interactions when naively thinking about the continuum theory --- let us consider an even simpler example: a massless Euclidean 2d Dirac fermion $\Psi$ with a
current-current Thirring interaction term,
\begin{align}
    S_{\Psi} = \int d^2{x}\, \left[ \bar{\Psi} \slashed{\partial} \Psi + 
    g_t (\bar{\Psi} \gamma^{\mu} \Psi )^2 \right]\,.
    \label{eq:thirring}
\end{align}
 This simple field theory has
the standard vector and axial symmetries $U(1)_V$ and $U(1)_A$, and the Thirring coupling $g_t$ is exactly
marginal and does not run\cite{Thirring:1958in}. After gauging
fermion parity $(-1)^F$ (which just means summing over periodic and anti-periodic conditions on e.g. a spacetime torus), it is known that the Thirring model gets a dual description as a
scalar field theory\cite{Coleman:1974bu,Coleman:1975pw,Coleman:1976uz}
\begin{align}
    S_{\varphi} = \int d^2{x}\, \frac{R^2}{4\pi} (\partial_{\mu} \varphi)^2
    \label{eq:varphi_action}
\end{align}
where $R^2 = \frac{1}{2} + \frac{g_t}{\pi}$, the scalar $\varphi$ is compact
$\varphi = \varphi + 2\pi$, and $j_V^{\mu} = \bar{\Psi} \gamma^{\mu} \Psi \sim i
\epsilon^{\mu\nu}\partial_{\nu} \varphi$, $j_A^{\mu} = \bar{\Psi}
\gamma^5\gamma^{\mu} \Psi \sim \partial^{\mu} \varphi $. The vanishing of the $\beta$ function for $g_t$ translates to parameter $R$ being marginal, which is indeed clear by inspection of the free-field action in Eq.~\eqref{eq:varphi_action}.

It is known that 2d compact scalars with $R$ and $1/R$ are T-dual to each other, in the
 sense that physics at $R$ is the same as at $1/R$ with $\varphi$ exchanged with
 its dual field $\theta$, related via $\partial_{\mu} \varphi = R^{-2}
 \epsilon_{\mu \nu} \partial^{\nu}\theta$.  The $T$-dual action is 
\begin{align}
    S_{\theta} = \int d^2{x}\, \frac{1}{4\pi R^2} (\partial_{\mu} \theta)^2 \,.
    \label{eq:theta_action}
\end{align}
The $2\pi$ periodic field $\theta$ transforms by shifts under $U(1)_V$, while
its winding current $i \epsilon^{\mu\nu}\partial_{\nu} \theta$ is identified
with $U(1)_A$.   

The `Dirac point' $R = \sqrt{1/2}$, which correcponds to $g_t=0$, enjoys a somewhat exotic enhanced $0$-form
symmetry.  In 2d, $0$-form symmetries are generated by topological line
operators. It is known that when $R = \sqrt{1/k}$ for $k\in\mathbb{Z}$,  $S_{\theta}$ is self-dual
under the combination of gauging $\mathbb{Z}_k \in U(1)_V$ and performing
T-duality.  As a result, there exists a topological line operator
$\mathcal{N}_k$ associated with performing these transformations in e.g. half of
spacetime\cite{Thorngren:2019iar, Thorngren:2021yso,Choi:2021kmx,Choi:2022zal}.  
The topological line operator $\mathcal{N}_k$ is a symmetry generator despite not
being not invertible for $k \neq 1$ (see e.g. Ref.~\cite{Shao:2023gho} for a nice review). This
means that no matter how we complicate the continuum model by e.g. adding
interactions, as long as $\mathbb{Z}_2 \in U(1)_V$ and $T$-duality remain unbroken, the line operator $\mathcal{N}_2$ will remain topological, and the point $R =
\sqrt{1/2}$ will be protected by the non-invertible symmetry
associated with $\mathcal{N}_2$.  This gives a symmetry-based way to understand why $g_t$ remains zero at all length scales if it is set to zero at short distances.

Can $R$ --- and hence $g_t$ --- get renormalized if the short-distance version of the model is defined
on a spacetime lattice? To understand why the answer is yes, let us discretize e.g.
$S_{\theta}$ using the Villain formalism
\begin{align}
    S_{\rm lat} = 
    \sum_{\ell} \frac{1}{4\pi R_{\rm lat}^2} 
    \left[(d\theta)_{\ell} - 2\pi n_{\ell}\right]^2 
    \label{eq:S_lat}
\end{align}
where $\theta_s \in \mathbb{R}$, $n_{\ell} \in \mathbb{Z}$, $(d\theta)_{\ell} =
\theta_{s+\hat{\ell}} - \theta_s$, $s, \ell$ denote sites and links, and $R_{\rm lat}$ is the lattice analog of $R$.  This action has a discrete gauge redundancy $
    \theta_s \to \theta_s + 2\pi k_s,\, n_{\ell} \to n_{\ell} + (dk)_{\ell}$
with $k_s \in \mathbb{Z}$, so that $n_{\ell}$ is a discrete gauge field
associated with $2\pi$ shifts of $\theta$. The $U(1)_V$ symmetry acts by constant shifts of $\theta_s$, but as is common in lattice discretizations of continuum field theories
with chiral symmetry, $U(1)_A$ is not preserved at finite lattice spacing.  To see this note that the chiral charge $Q_A$ within a
region bounded by a closed curve $C$ is the winding number of $\theta$,
\begin{align}
    Q_A(C) = -  \frac{1}{2\pi} \sum_{\ell \in C} \left[(d\theta)_{\ell} - 2\pi n_{\ell}\right]
     \in \mathbb{Z} \,.
\end{align} 
Unfortunately the path integral involves  sums over  all possible $n_{\ell}$, and so $Q_A$ is not conserved with this lattice discretization. 

We can now understand why the point  $R_{\rm lat} = \sqrt{1/2}$ does not have any
enhanced symmetry on the lattice.  The appearance of an enhanced symmetry at $R = \sqrt{1/2}$ in the
continuum model required both $U(1)_V$ symmetry and T-duality.  But while the
lattice model in Eq.~\eqref{eq:S_lat} preserves $U(1)_V$, it does not preserve
T-duality: for small $R_{\rm lat}$ the model is gapped while for large $R_{\rm lat}$ the physics is completely different and the model is gapless, see e.g.\cite{Kosterlitz:1973xp,Janke:1993va}.   Given that $R_{\rm lat} = \sqrt{1/2}$ does not have any enhanced symmetries, $R_{\rm lat}$ should get
renormalized as we go to the continuum limit. Reaching a continuum limit
with $g_t = 0$ should require tuning $R_{\rm lat}$ to a bare value which is \emph{not} $R_{\rm
lat} = \sqrt{1/2}$.   

That this renormalization really does occur can be read off from numerical and
analytic lattice calculations\cite{Kosterlitz:1973xp,Janke:1993va}. A continuum analysis based on Eq.~\eqref{eq:theta_action} implies
that `vortex operators' $e^{i\theta}$ become relevant when $R$ hits the enhanced-symmetry value $2$. If the relation $R_{\rm lat} = R$ were to hold, then when $R_{\rm lat}$ goes through
$2$, the system should go through a BKT transition, from a conformal field
theory phase with an emergent $U(1)_A$ symmetry to a gapped phase with no
$U(1)_A$ symmetry. However, the critical value of $R_{\rm lat}$ is known to be
\cite{Kosterlitz:1973xp,Janke:1993va}
\begin{align}
    R_{\rm lat} \approx 2.16 > 2 \,.
\end{align}
So $R_{\rm lat}$ --- and hence also $(g_t)_{\rm lat}$ --- is indeed renormalized in the lattice theory, despite what one might have expected when considering the continuum theory.

This section illustrates the familiar lesson that parameters can get renormalized differently in lattice versus continuum field theories. To prevent a parameter from being renormalized, it must be a point of enhanced symmetry in whichever theory one considers, be it in the continuum or on the lattice. In particular, when studying QCD[Adj] using lattice Monte Carlo methods one should interpret the bare values of $\lambda_j$ and $\lambda_m$ as part of its parameter space, and studying any particular point in the physical $(\lambda_j,\lambda_m)$ plane is likely to require fine-tuning of the bare parameters.

\section{Conclusions}
\label{sec:conclusions}
Two-dimensional  QCD with one massless adjoint fermion is an interesting setting
 for studying confinement, with surprisingly rich and subtle physics.  These
 subtleties include a rich set of symmetries and anomalies, including ones that
 involve an exotic non-invertible symmetry\cite{Komargodski:2020mxz}. These
 anomalies can have the surprising result of driving 2d QCD[Adj] into a deconfined
 phase characterized by a spontaneously broken $\mathbb{Z}_N$ $1$-form
 symmetry\cite{Dalley:1992yy,Kutasov:1993gq,Bhanot:1993xp,Gross:1997mx,
 Komargodski:2020mxz,Delmastro:2021otj,Delmastro:2022prj}.
 Another interesting subtlety is the possibility of adding two classically
 marginal four-fermion interaction terms to the action of the theory for $N>2$.

In this paper we have explored the interplay of these subtleties by viewing
QCD[Adj] as a Wilsonian effective field theory, where the dimensionless
four-fermion couplings $\lambda_j, \lambda_m$ must be interpreted as part of its
parameter space.    These couplings can easily be produced by short-distance
physics such as a lattice regularization, as we have discussed in
Section~\ref{sec:lattice}.  Even if they appear with small (positive) values,
these couplings turn out to run and generically become large at long distances,
as discussed in Section~\ref{sec:beta_functions}.  Therefore 2d adjoint QCD can
be viewed  as a theory with \emph{three} dimensionful parameters: the 't Hooft
coupling $\lambda$ and the strong scales $\Lambda_j$ and $\Lambda_m$ associated
with $\lambda_j, \lambda_m$\cite{Cherman:2019hbq}, rather than just one
dimensionful coupling $\lambda$, as was done historically.

Our results are likely to be helpful for lattice Monte Carlo studies of 2d
QCD[Adj] on spacetime lattices.  Without careful fine tuning, these numerical
simulations are likely to probe the physics of adjoint QCD with non-zero values
of $\lambda_j$ and $\lambda_m$. One way to check whether e.g. the $\lambda_m$
coupling is generated in a lattice simulation is to calculate
\begin{align}
    \langle \tr \bar{\psi} \psi \tr \bar{\psi} \psi \rangle \,
    \label{eq:lambda_m_VEV}
\end{align}
at finite volume.  This expectation value must vanish in a QFT that enjoys the
non-invertible symmetry uncovered in Ref.~\cite{Komargodski:2020mxz}.  But if $\lambda_m = 0$ in the bare lattice action while $ \langle \tr \bar{\psi}
\psi \tr \bar{\psi} \psi \rangle  \neq 0$  even in the chiral limit, then one
can conclude that the non-invertible symmetry is explicitly broken and the
$\lambda_m$ term has been radiatively generated.  One can quantify the size and
sign of the radiatively-generated $\lambda_m$ by turning on a bare $\lambda_m$
coupling and tuning it to get \eqref{eq:lambda_m_VEV} to vanish.  However, as we
argued in Section~\ref{sec:lattice}, it is natural to expect the coefficients of
the four-fermi terms to be small near
the continuum limit, scaling as $\sim (ga)^{p},\, p>1$.   Therefore it is likely to be easiest to examine
\eqref{eq:lambda_m_VEV} on small lattices with $ga \sim
1$, i.e. far from the continuum limit.  After that one could explore the behavior of \eqref{eq:lambda_m_VEV} near
the continuum limit in large boxes with characteristic size $L \gg g^{-1}$.

The fact that $\lambda_j$ and $\lambda_m$ can get large in the long-distance
limit implies that  e.g. the continuum-limit particle spectra produced from
spacetime lattice Monte Carlo simulations may differ appreciably from the
particle spectra that have been extracted from light-cone calculations with
$\lambda_j = \lambda_m = 0$, unless the Monte-Carlo simulations are carefully
fine-tuned.  In future work it would be interesting to explore whether e.g.
domain-wall or overlap fermion lattice actions induce $\lambda_m \neq 0$ and
$\lambda_j \neq 0$ as one goes to the continuum limit, as well as to numerically
explore what happens to the spectrum as one dials $\lambda/\Lambda_j$ and
$\lambda/\Lambda_m$.

{\bf Acknowledgements.}. We are grateful to O. Aharony, D. Delmastro, D.
Gaiotto, I. Klebanov, Z. Liu, P. Shanahan, and M. Strassler for helpful
discussions, and A. C. thanks T. Jacobson, Y. Tanizaki, and M. \"Unsal for
kindling his interest in the topic of this work.  This work was supported in
part by Simons Foundation through the Collaboration on Confinement and QCD
Strings under award number 994302 (A. C.) and by the National Science Foundation
Graduate Research Fellowship under Grant No. 1842400 (M. N.).

\bibliography{main}

\end{document}